\documentstyle[preprint,aps,graphicx,floats]{revtex}

\begin{document}

\newcommand{\lbl}[1]{\label{eq:#1}}
\newcommand{ \rf}[1]{(\ref{eq:#1})}
\newcommand{\be}{\begin{equation}}
\newcommand{\en}{\end{equation}}
\newcommand{\bea}{\begin{eqnarray}}
\newcommand{\ena}{\end{eqnarray}}
\newcommand{\lapprox}{%
\mathrel{%
\setbox0=\hbox{$<$}
%\setbox1=\hbox{$\sim$}
\raise0.6ex\copy0\kern-\wd0
\lower0.65ex\hbox{$\sim$}
}}
\newcommand{\gapprox}{%
\mathrel{%
\setbox0=\hbox{$>$}
%\setbox1=\hbox{$\sim$}
\raise0.6ex\copy0\kern-\wd0
\lower0.65ex\hbox{$\sim$}
}}

\newcommand{\tr}{\mbox{\rm tr}}
\def\hu{\hat{U}}
\def\hc{\hat{\chi}}
\def\nf{$N_f$}
\def\nc{$N_c$}

\title{
\[ \vspace{-2cm} \]
\noindent\hfill\hbox{\rm  DFPD-01/TH/08} \vskip 1pt
\noindent\hfill\hbox{\rm  IPN-DR/01/005} \vskip 10pt
$\eta^\prime$ Mass and Chiral Symmetry Breaking
at Large $N_c$ and $N_f$
}  
\author{
L.~Girlanda${}^a$, J. Stern${}^b$ and P. Talavera${}^b$
}

\address{
$^a$ Dipartimento di Fisica ``Galileo Galilei'',
Universit\`a di Padova and INFN, 
via Marzolo 8, I-35131 Padova, Italy.\\
$^b$ Groupe de Physique Th\'eorique,
Institut de Physique Nucl\'eaire,
Universit\'e Paris Sud,
F-91406 Orsay Cedex, France} 
\maketitle

\begin{abstract}
We propose a method for implementing the large-$N_c$, large-$N_f$ limit of
QCD at the effective Lagrangian level. Depending on the value of the ratio
$N_f/N_c$, different patterns of chiral symmetry breaking can arise, leading
in particular to different behaviors of the $\eta^\prime$-mass in the
combined large-$N$ limit.
%\\ {\bf Packs:} 11.15.Pg, 12.39.Fe, 12.38.Aw
\end{abstract}
\pacs{11.15.Pg, 12.39.Fe, 12.38.Aw}
%\vskip2pc]

{\bf 1.}~Large-\nc~considerations successfully explain many non-perturbative 
aspects of confining gauge theories \cite{tH,witten}. 
However, there are
at least two exceptions -both related to a strong OZI rule violation- in
which the $1/N_c$ expansion apparently fails:
$(i)$~In the scalar channel the spectrum is not dominated by a nonet
of ideally mixed states and chiral symmetry breaking exhibits an
important dependence on the number \nf~ of light quark flavors \cite{orsay};
$(ii)$ At large~\nc, the $\eta^\prime$-field
becomes massless due to its relation
to the U(1) anomaly while Nature realizes it
like a heavy state. 
In this note we reconsider these problems
in the limit in which
both \nf\, and \nc\, tend to infinity with fixed ratio
\cite{Veneziano}. Since, at least at
lowest orders of perturbation theory, the (rescaled) QCD
$\beta$-function (with $g^2 N_c\equiv$ const.)
only depends on the ratio $N_f/N_c$, we
might expect that the hadronic spectrum resembles the physical one, in
particular with chiral symmetry breakdown and $\Lambda_H \sim 1$~GeV.
On the other hand, several hints ({\em e.g.} from the study of the conformal window, in
QCD and its supersymmetric version), suggest a non trivial phase structure
of the theory as a function of \nf\, and \nc. One can in principle distinguish
three different phases, characterized by different symmetries of the vacuum,
depending on the ratio \nf/\nc: $(a)$~for low \nf/\nc\, only the SU$_V$(\nf)
remains unbroken; $(b)$~for higher \nf/\nc\, the vacuum is invariant under a
larger group, SU$_V$(\nf) $\times$ Z$_{\mathrm{chiral}}$(\nf), where
Z$_{\mathrm{chiral}}$(\nf) is the center of the chiral symmetry group
SU$_L$(\nf)$\times$SU$_R$(\nf) \cite{Dashen:1969eg,kks};
$(c)$~ for high \nf/\nc\, no spontaneous
symmetry breaking takes place (and hence no confinement) and the symmetry of
the vacuum is the whole SU$_L$(\nf)$\times$SU$_R$(\nf).
Notice that case $(b)$ corresponds
to the maximal possible symmetry of the vacuum
 in a confining vector-like theory.
The existence of this phase is an assumption related
to
the issue of the non-perturbative renormalization
of the bare Weingarten's inequalities comparing axial-axial and
pseudoscalar-pseudoscalar two point functions \cite{kks,wein}.
We model the combined large-\nf, large-\nc\, limit by adding to the usual light
flavors $q=(u,d,s)$, a set of $N$ auxiliary flavors $Q=(Q_1,\ldots ,Q_N)$ of
common mass $M \gg m_q$, but still $M\ll\Lambda_H$. This mass $M$ should be
considered sufficiently  small so that a power series expansion makes sense,
but  simultaneously much larger than any of the light quark masses $m_q$,
thus the  auxiliary fields can be integrated out at sufficiently low-energy.
We then formally deal with $N_f=N+3 \equiv n \to \infty$ flavors, but only
the three lightest ones are physical.
The r\^ole of these
auxiliary flavors should be analogous to the one of the strange quark, when
one considers the SU(2)$\times$SU(2) chiral dynamics of $u$ and $d$ quarks.

\indent

{\bf 2.}~Let \nf/\nc\, be subcritical, so that we are in the
Z$_{\mathrm{chiral}}(n)$-asymmetric phase. The $n^2-1$ (pseudo) 
Goldstone bosons (GB)
can be collected in a matrix $\hat U(x) \in\, $SU($n$), (hereafter $n\times n$
matrices will be denoted by a hat) and their 
low-energy dynamics 
can be described by the effective Lagrangian
\begin{equation} \label{eq:sublag}
{\cal L}_{\mbox{\rm \tiny{sub}}} = 
\frac{F^2}{4} \Big\{
\langle D_\mu \hu D^\mu \hu^\dagger \rangle
+2 B_0 \langle \hu^\dagger \hc + \hc^\dagger \hu \rangle
\Big\}\,.
\end{equation}
where $\hc$ is the scalar-pseudoscalar source,
\begin{equation}
{\cal L}^{\mathrm{QCD}}_{\hc} = - \bar \Psi_L \hc \Psi_R - \bar \Psi_R
\hc^{\dagger} \Psi_L, \quad \Psi=\left(\begin{array}{c} q \\ Q
\end{array} \right)\,,
\end{equation}
and $\langle \ldots \rangle$ denotes flavor trace.
The $n\times n$ source matrix will be chosen as
\begin{equation} \label{eq:chi-n3}
\hc = \left( \begin{array}{cc} \chi & {\bf 0} \\
{\bf 0} & M {\mathrm{e}}^{i \theta/N} {\bf 1}_{N\times N} \end{array}
\right)\,,
\end{equation}
where $\chi$ is the $3\times3$ light quark source (mass term), $\theta$ is
the vacuum angle and $M$ is real and positive. 
Notice that there are no sources attached to the $N$ auxiliary flavors and the 
corresponding GB degrees of freedom are frozen: in the tree
approximation the $n\times n$ GB field matrix becomes
\begin{equation} \label{eq:u-n3}
\hu = \left( \begin{array}{cc} U {\mathrm{e}}^{i \varphi/3}  & {\bf 0} \\
{\bf 0} & {\mathrm{e}}^{-i \varphi /N} {\bf 1}_{N\times N} \end{array} \right)
\in {\mathrm{SU}(n)}\,,
\end{equation}
where $U \in$ SU(3) collects the eight physical GB fields.
The remaining U(1) field $\varphi$ will be interpreted as the
$\eta^{\prime}$-field: for $\chi= m {\bf 1}_{3\times3}$ (no mixing),
\begin{equation}
\eta^{\prime} = F \sqrt{\frac{n}{6N}} \varphi
\end{equation}
and the corresponding mass, for $m=0$ is
\begin{equation}
M^2_{\eta^{\prime}} = \frac{6 B_0 M}{n} \to 0\,,
\end{equation}
which vanishes in the (combined) large-$N$ limit. Hence, {\em in this
phase}, $\eta^{\prime}$-mass behaves as in the standard $N_c \to \infty$,
$N_f$-fixed limit.

\indent

\begin{figure}
\leavevmode
\begin{center}
\includegraphics[width=10cm]{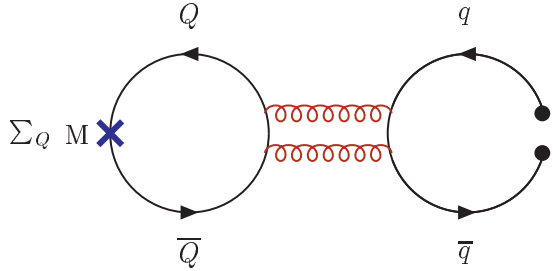}
\end{center}
\caption{\label{fig:condensate} 
  Representation of the contribution to the induced condensate
  Eq.~(\ref{eq:res2}). The cross refers to $\overline{Q}Q$
  insertion.
  }
\end{figure}

{\bf 3.}~{F}or higher \nf/\nc\, we expect the
Z$_{\mathrm{chiral}}$($n$)-symmetric phase to occur: the vacuum is
symmetric under 
\begin{equation}
\Psi_{L,R} \to {\mathrm{e}}^{2 \pi i \frac{k_{L,R}}{n}} \Psi_{L,R},
\quad k_{L,R} =1,\ldots ,n-1\,,
\end{equation}
in addition to the usual SU$_V$($n$). Notice that this
Z$_{\mathrm{chiral}}$($n$) is also a subgroup of the (anomalous)
U$_L$(1)$\times$U$_R$(1). This additional symmetry of the vacuum finds its
natural interpretation within the effective theory described by the
Lagrangian ${\cal L} ( \hu, \hc, \theta)$. The GB field $\hu(x)\in$\,SU($n$) is
usually understood as simply connected to~${\bf 1}$: $\hu (x) = {\bf 1} + i
\varphi_a(x) T^a + \ldots $ and, in the corresponding effective theory, the
integration measure $ {{\cal D}} \hu $  is treated accordingly. However SU($n$) is
{\em not} simply connected. We are free to choose an integration measure
treating all sectors of SU($n$) alike:
\begin{equation}
\label{eq:moduli}
\hspace{-0.2cm}
\int \!\! {\cal D} \hu  {\mathrm{e}}^{i \int dx {\cal L}(\hu,\hc,\theta)} 
\hspace{-0.1cm} 
\to
\hspace{-0.1cm} 
\int \!\! {\cal D} \hu  \sum_{k=0}^{n-1} {\mathrm{e}}^{i \int dx {\cal L}(\hu
{\mathrm{e}}^{-2 \pi i k/n},\hc,\theta)} \,,
\end{equation}
where $\hu$ and ${\cal D} \hu$ again concern the connected vicinity of~${\bf 1}$. This freedom derives from the fact that an effective theory is
merely constrained  by Ward identities (WI), which fix its {\em local} but not
its
global aspects. In particular the usual solution of the anomalous U(1) 
WI only guarantees 
${\cal L} (\hu,\hc,\theta) = {\cal L}(\hu,\hc {\mathrm{e}}^{i
\theta/n})$ for $\theta \sim 0$. However, under the
 Z$_{\mathrm{chiral}}$($n$) transformation one has
\begin{equation}
 {\cal L} (\hu {\mathrm{e}}^{- \frac{2 \pi i k}{n}},\hc {\mathrm{e}}^{i
\theta/n}) = {\cal L} (\hu ,\hc {\mathrm{e}}^{i \frac{
\theta + 2 \pi k}{n}})\,,
\end{equation}
i.e. the above prescription~(\ref{eq:moduli}) {\em restores the $2
\pi$-periodicity} in the vacuum angle \cite{hz}.
The Z$_{\mathrm{chiral}}$($n$)-symmetry of the vacuum expressed in terms
of a local Lagrangian amounts to the constraint
\begin{equation}
 {\cal L} (\hu {\mathrm{e}}^{- \frac{2 \pi i k}{n}},\hc,\theta) = {\cal L} (\hu
,\hc,\theta) = {\cal L} (\hu ,\hc {\mathrm{e}}^{i \frac{\theta+ 2 \pi k}{n}})\,,
\end{equation}
which is not necessarily true in
general and it will be taken as a \emph{definition} of the 
Z$_{\mathrm{chiral}}$($n$)-symmetric phase. 
The effective Lagrangian exhibiting  Z$_{\mathrm{chiral}}$($n$)-symmetry
consists of two parts, ${\cal L}= {\cal L}_1 + {\cal L}_2$. ${\cal
L}_1$ has the whole continuous U$_A$(1) symmetry, $\hc \to {\mathrm{e}}^{i
\alpha} \hc$, i.e. it is independent of the $\theta$ angle,
\begin{equation}
\label{eq:treelag}
{\cal L}_1 = \frac{F^2}{4} \Big\{
\langle D_\mu \hu D^\mu \hu^\dagger \rangle
+Z \langle \hu^\dagger \hc \rangle \langle \hc^\dagger \hu \rangle
+\ldots 
\Big\}\,,
\end{equation}
where dots stand for pure source and higher orders terms.
${\cal L}_2$ consists of terms which break U$_A$(1) down to 
Z$_{\mathrm{chiral}}$($n$). The lowest order term with this property reads
\begin{eqnarray}
\label{eq:holo}
&&{\cal L}_2  = \hspace{-0.1cm}
\! \sum_k^{n-1} \hspace{-0.2cm} \sum_{\{j_1\ldots j_k\}} \hspace{-0.2cm}
A^{(k)}_{\{j_1\ldots j_k\}}
\langle \left( \hu^\dagger \hc \right)^{j_1}\rangle   \ldots
\langle \left( \hu^\dagger \hc \right)^{j_k}\rangle
 + {\mbox{\rm h.c}}\,, \nonumber\\ &&
{\mbox{\rm where}} \quad 
j_1,\ldots ,j_k = 1,\ldots ,n,\quad j_1 + \ldots +j_k=n\,.
\end{eqnarray}
Despite the fact that the Lagrangians~(\ref{eq:treelag}) and~(\ref{eq:holo})
are of different orders in $\hc$ they can coexist, since they describe two
independent sectors of the effective theory: Eq.~(\ref{eq:holo}) is {\em
holomorphic} and is not renormalized by loops arising from the U(1)-invariant
sector as represented by Eq.~(\ref{eq:treelag}).
Eq.~(\ref{eq:holo}) becomes more transparent using  Eqs.~(\ref{eq:chi-n3})-(\ref{eq:u-n3})  and expanding in powers of
the light quark masses $\chi$. Denoting by $W$ the $3\times3$ matrix and by
$\zeta$ the phase factor,
\begin{equation} \label{eq:Wzeta}
W=U^{\dagger} \chi {\mathrm{e}}^{- i \frac{\varphi}{3}}, \quad \zeta=
{\mathrm{e}}^{i \frac{\theta + \varphi}{N}}\,,
\end{equation}
Eq.~(\ref{eq:holo}) can be rewritten as
\begin{eqnarray} 
\label{eq:second}
&&
{\cal L}_2  = 
a_n ( M \zeta )^n + b_n \langle W \rangle (M \zeta )^{n-1} + c_n \langle W^2
\rangle (M \zeta)^{n-2} \nonumber \\
&&+ d_n \langle W \rangle^2 (M \zeta)^{n-2} + {\mathrm{h.c.}} + {\cal
O}(W^3)\,.
\end{eqnarray}
Similarly, the reduction SU($n$)$\to$SU(3)$\times$U(1) of the component ${\cal
L}_1$ can be expressed in terms of the variables~(\ref{eq:Wzeta}) as
\begin{eqnarray}
{\cal L}_1 &=& \frac{F^2}{4} \biggl\{ \langle D_{\mu} U^{\dagger} D^{\mu} U
\rangle + \frac{n}{3 N} \partial_{\mu} \varphi \partial^{\mu} \varphi
\nonumber \\
&& + M N Z \langle W \zeta^{\dagger} + W^{\dagger} \zeta \rangle + Z \langle
W^{\dagger} \rangle \langle W \rangle \biggr\} + \ldots \,.
\end{eqnarray}
In the tree approximation the $\eta^\prime$-mass merely arises from the
holomorphic part ${\cal L}_2$, ($m_u=m_d=m_s=0$),
\begin{equation} \label{eq:res1}
M^2_{\eta^{\prime}} = \frac{12 N}{n}  \frac{1}{F^2} a_n M^n\,.
\end{equation}
In contrast to the Z$_{\mathrm{chiral}}$($n$)-asymmetric phase, [cf. Eq.~(\ref{eq:sublag})],
which contains a genuine condensate term $B_0$, such a term is absent in the
Z$_{\mathrm{chiral}}$($n$)-symmetric phase. However, in the
reduction~(\ref{eq:chi-n3})-(\ref{eq:u-n3}), an {\em induced condensate}
appears through
the OZI violation terms [see Fig.~(\ref{fig:condensate})]
(the Z$_{\mathrm{chiral}}$($n$)-symmetry is explicitly broken by the $Q$-mass term),
\begin{equation} \label{eq:res2}
F^2 B_{\mathrm{induced}} = \frac{1}{2} F^2 M N Z + 2 b_n M^{n-1}\,,
\end{equation}
where the first term on the r.h.s. arises from ${\cal L}_1$, whereas the
second term comes  from the holomorphic Lagrangian. The quadratic terms in
$W$ of Eq.~(\ref{eq:second}) contribute to the subleading low-energy
constants $L_6$, $L_7$, $L_8$ \cite{gl},
denoted by hat
\begin{equation} \label{eq:res3}
\begin{array}{lcl}
B^2_{\mathrm{induced}} (\hat L_6 - \hat L_7 ) & = & \frac{1}{32} F^2 Z\,, \\
B^2_{\mathrm{induced}} \hat L_8 & = & \frac{1}{4} c_n M^{n-2}\,, \\
B^2_{\mathrm{induced}} (\hat L_6 + \hat L_7 ) & = & \frac{1}{4} d_n M^{n-2} \,.
\end{array}
\end{equation}

\indent

{\bf 4.}~We now turn to the leading behavior of all these induced low-energy
constants in the combined large-$N$ limit.
We consider (connected) correlators of quark bilinears $\bar \Psi \Gamma
\Psi$. Usual large-\nc\, counting rules are maintained. In addition, every
quark loop gives rise to a {\em flavor trace} involving all flavor matrices 
contained in that loop. Consequently, each internal (``sea'') quark loop
will be {\em enhanced} by a 
factor $N_f=n$ and suppressed (as usual) by a factor $1/N_c$.
(However, if a quark loop is {\em valence}, i.e.
attached to an external flavor source, it will not lead to a flavor-enhancement
factor.)
In particular, the constants $F^2$ and $Z$ in
Eq.~(\ref{eq:treelag}) behave as $F^2 \sim N$ and $Z\sim 1/N$.
The contribution to the fermionic determinant can be formally written 
(in a large Euclidean box) like
\begin{equation}
\Delta = \exp \sum_k \log 
\left( 1+ \frac{ \lambda_k^2 -\omega_k^2}{\omega_k^2 + M^2}
\right)^n\,,
\end{equation}
where $\lambda_k$ are  
Dirac operator eigenvalues and $\omega_k$ the corresponding eigenvalues in
the absence of interactions. 
Hence at large-\nc\, one
expects 
$\lambda_k^2 - \omega_k^2 \sim {\cal O}(g^2) \sim {\cal O}(1/N_c)$. 
This illustrates the mechanism by which the fermionic determinant stays non
trivial and finite in the combined large-$N$ limit,
merely depending on the ratio \nf/\nc.
The large-$N$ counting of the holomorphic part is more subtle: we deal with
a large-$N_f$, large-\nc\, behavior of a large-$N$-point function. In the tree
approximation, the holomorphic part of the Lagrangian (at $\hu={\bf 1}$) 
is connected 
with a QCD correlation function
in a Euclidean finite volume $V$
\begin{eqnarray} \label{eq:match}
- {\cal L}_2({\bf 1},\hc) &=&-\sum_k^{n-1} \sum_{\{j_1\ldots j_k\}}
A^{(k)}_{\{j_1\ldots j_k\}}
\langle   \hc^{j_1}\rangle   \ldots
\langle \hc^{j_k}\rangle \nonumber \\
& =& \frac{1}{n!} \frac{1}{V} \langle \left[ \int dx \bar \Psi_L \hc \Psi_R(x)
\right]^n \rangle_{\mathrm{con}}\,.
\end{eqnarray}
The average on r.h.s. of Eq.~(\ref{eq:match}) can be evaluated
 at non zero masses
$m$ and $M$. The other chirality part $\bar \Psi_R \Psi_L$ will contribute but
{\em not} to the lowest order $\hc^n$ of the holomorphic part of the
Lagrangian. Let us introduce the notation
\begin{equation}
\begin{array}{l}
K= \int dx \sum_{i=1}^N \bar Q_L^i(x)\, Q_R^i(x)\,, \\
k_i =\int dx \, \bar q_{i L}(x)\, q_{i R}(x)\,, \quad i=1,2,3\,.
\end{array}
\end{equation}
Choosing in Eq.~(\ref{eq:match}) 
$\hc={\mbox{\rm diag}}(m_1,m_2,m_3,M,\ldots ,M)$, and combining it with
 Eq.~(\ref{eq:second}) one gets
\begin{eqnarray} \label{eq:abcd}
&&- \left\{a_n M^n + b_n M^{n-1} (m_1 + m_2 + m_3) \right. \nonumber \\
&& + \left[ c_n  (m_1^2 + m_2^2 +
m_3^2) + d_n (m_1 + m_2 + m_3)^2 \right] M^{n-2} \nonumber \\ 
&&+\ldots \left. \right\} 
=\frac{1}{n!} \frac{1}{V} \langle (M K + m_1 k_1 + m_2 k_2 + m_3
k_3)^n \rangle_{\mathrm{con}}\,.
\end{eqnarray}
Comparing the coefficients of $M^n$
\begin{equation}
\label{eq:an}
a_n M^n = -\frac{1}{n!} M^n \frac{1}{V} \langle K^n \rangle_{\mathrm{con}}\,.
\end{equation}
The single quark loop contribution to Eq.~(\ref{eq:an}) reads
\begin{eqnarray}
\label{eq:QCD}
&&\frac{1}{V} \langle K^n\, \rangle = - (n-1)! \frac{1}{V} \langle \langle
\int dx_1 \ldots  dx_n {\mathrm{Tr}} \biggl\{ S^{RL}(x_1,x_2) \nonumber \\
&& S^{RL}(x_2,x_3)\ldots S^{RL}(x_n,x_1) \biggr\}
\rangle \rangle {\mathrm{Tr}}({\bf 1}_{N\times N})\,, 
\end{eqnarray}
where $S^{RL}(x,y)$ denotes the chiral part of the fermion propagator
\[
S^{RL}(x,y) \hspace{-0.1cm}
= \hspace{-0.1cm} \left( \frac{1 + \gamma_5}{\sqrt{2}} \right) \hspace{-0.2cm}
\sum_{\lambda_k \ge 0} \frac{M}{M^2 + \lambda_k^2} \varphi_k(x)
\varphi_k^\dagger(y) \left(\frac{1 + \gamma_5}{\sqrt{2}}\right)
\]
in terms of the orthonormal Fujikawa chiral
basis \cite{fuji} ($\varphi_k$ is the Dirac
eigenvector belonging to the eigenvalue $\lambda_k$)
 and  $\langle \langle \ldots  \rangle \rangle$ stands for the
average over gluon configurations with insertion of fermionic determinant.
The 
factor $(n-1)!$ in Eq.~(\ref{eq:QCD}) counts the different ways
of connecting $n$ points by a single {\em one quark loop}.
In fact a closer 
examination of the combinatorics 
of multiloop diagrams' contributions to $\langle K^n \rangle$ in 
Eq.~(\ref{eq:QCD}) reveals that none of them is more important
that 
the one with the least number of quark loops.
The integrals in Eq.~(\ref{eq:QCD}) can be performed with the result
\begin{equation}
\label{an}
a_n M^n \sim \lim_{ \stackrel{ \mbox{\scriptsize$V\rightarrow\infty$} }
{n\rightarrow\infty} } 
\frac{1}{V} \langle \langle \sum_{\lambda_k \ge 0} \left( 1
+ \frac{\lambda_k^2}{M^2} \right)^{-n} \rangle \rangle\,.
\end{equation}
Even if we do not consider here the chiral limit
$M\rightarrow 0$, the behavior of Eq.~(\ref{an}) is merely controlled
by the average density of small Dirac eigenvalues.
Indeed, for any fixed $M$ and (arbitrary small) $\epsilon$ the 
eigenvalues
$\lambda_k^2 \geq \epsilon$ do not contribute to the large-$N$ limit of
Eq.~(\ref{an}). The latter should be of the order
of the average number of states ${\cal N}_\epsilon$~with 
$\lambda_k^2 \leq \epsilon$.
On general grounds one expects ${\cal N}_\epsilon 
\sim V N_c$ \cite{lesm}. (The density
of states should grow proportionally with $N_c$.)
Hence, in the combined large-$N$ limit
$a_n M^n \sim {\cal O}(N_c)$ and according to Eq.~(\ref{eq:res1})
\begin{equation}
\label{eq:etamass}
M_{\eta^\prime}^2 \sim {\mathrm{const}}\,,
\end{equation}
thus not suppressed anymore. Similar conclusions have been reached also
in different contexts \cite{Hsu,Minkowski:1998hv}.
The remaining coefficients in Eq.~(\ref{eq:abcd}) 
can be found similarly:
\begin{eqnarray}
&&b_n M^{n-1}\sim - \frac{M^{n-1}}{(n-1)!} \frac{1}{V} \langle K^{n-1} k_1
\rangle_{\mathrm{con}}\,, 
 \\&&
(c_n + d_n)M^{n-2} \sim - 
\frac{M^{n-2}}{(n-2)!} \frac{1}{V} \langle K^{n-2} k_1^2
\rangle_{\mathrm{con}}\,, 
\end{eqnarray}
receiving leading contribution from at least two quark loops and 
consequently suppressed by $1/N_c$ relative to Eq.~(\ref{an}). 
Finally 
\begin{equation}
d_n M^{n-2} \sim -\frac{M^{n-2}}{(n-2)!} \frac{1}{V} \langle K^{n-2} k_1 k_2
\rangle_{\mathrm{con}}\,,
\end{equation}
which involves at least three quark loops.
This leads to the final estimate
\begin{equation} 
\hspace{-0.2cm} b_n M^{n-1} \hspace{-0.1cm} \sim  c_n M^{n-2} 
\hspace{-0.1cm} \sim {\cal O}(1)\,,  \quad
d_n M^{n-2} \hspace{-0.1cm} \sim  {\cal O}(1/N_c)\,.
\end{equation}
As a consequence, the holomorphic contribution to the induced condensate, 
Eq.~(\ref{eq:res2}), is suppressed relative to the non-holomorphic one.
The latter is given by the OZI rule violating constant $Z$ which is 
suppressed by 1/\nc~ but this suppression is compensated by a flavor 
enhancement factor $N=n-3$. As a result
\begin{equation}
F^2 B_{\mathrm{induced}} \sim {\cal O}(N) + {\cal O}(1)\,,
\end{equation}
where the first term is the non-holomorphic and the second one the
holomorphic contribution.
{F}or the tree contribution quoted in Eq.~(\ref{eq:res3})
 the large-$N$ counting reads
\begin{eqnarray} \label{eq:LUN}
%\begin{equation} \label{eq:LUN}
%\begin{array}{lcl}
&&B^2_{\mathrm{induced}}\, (\hat L_6 - \hat L_7 )  \sim  
B^2_{\mathrm{induced}}\, \hat L_8  \sim  {\cal O}(1)\,, \nonumber \\ &&
B^2_{\mathrm{induced}}\, (\hat L_6 + \hat L_7 )  \sim  {\cal O}(1/N)\,.
%\end{array}
%\end{equation}
\end{eqnarray}

\indent

{\bf 5.}~More comments on the large-$N$ behavior of the $\eta^\prime$-mass
are in order. 
The usual argument for finding the behavior of the $\eta^\prime$-mass
\cite{witten} derives from the necessity to cancel the 
$\theta$-dependence of the pure gluodynamics when massless quarks
are added. This is only possible when the 1/\nc~ suppression of the internal 
quark loops is compensated by the $\eta^\prime$-pole contribution 
$M_{\eta^\prime}^{-2}$. This leads to the Veneziano-Witten's formula and the 
vanishing of $M_{\eta^\prime}^2$ as 1/\nc. 
On the other hand, in the combined
large-$N$ limit internal quark loops are not suppressed, and it is not
possible to isolate pure glue contributions by large-$N$ arguments. As a 
consequence the $\eta^\prime$-mass does not vanish anymore and its relation 
to the topological susceptibility is lost.
We may as well consider the $\eta^\prime$-field as heavy and integrate it out.
At tree level this amounts to the shift in the constant $L_7$ 
\begin{equation}
L_7 = \hat L_7 - \frac{F^2}{48 M_{\eta^\prime}^2}, \quad L_i = \hat
L_i, \quad (i \neq 7)\,,
\end{equation}
as it is seen by evaluating 
the singlet minus octet pseudoscalar two-point function
\cite{Leutwyler:1990pn}.

\indent

{\bf 6.}~Let us summarize the results of this work. We have asked
whether the combined large-$N$ limit (as defined in this paper) helps
understanding the peculiar properties of QCD in the vacuum and
$\eta^\prime$-channels. $(i)$ In the scalar channel, this limit suggests
a flavor enhancement of the OZI rule violation, leading in particular
to the emergence of an \emph{induced quark condensate} [Eq.~(\ref{eq:res2})]
on top of the genuine condensate $B_0$ [c.f. Eq.~(\ref{eq:sublag})].
In the large \nf/\nc~phase in which the genuine condensate is forbidden
due to the Z$_{\mathrm{chiral}}$($n$)-symmetry, the induced condensate plays
the r\^ole of $B_0$ in describing chiral symmetry breaking.
An induced condensate would manifest itself by an important flavor
dependence [as in
Eq.~(\ref{eq:res2})] and it could be, in principle, disentangled from
$B_0$ in this way \cite{orsay}. $(ii)$ A distinctive feature of the
Z$_{\mathrm{chiral}}$($n$)-symmetric phase is the non-vanishing
$\eta^\prime$-mass in the combined large-$N$ limit. (Since this phase is expected for large \nf/\nc, the usual large-$N_c$, fixed-$N_f$ arguments do not
apply.) In particular, the relation of the $M_{\eta^\prime}$ to the axial
anomaly and to the topological susceptibility of the pure YM theory are now
modified. $(iii)$ The place of the scale $M$ in building $\eta^\prime$-mass, 
induced condensate and the low-energy constants $L_6,L_7$ and $L_8$ is important
and not entirely understood. Special attention should be payed to the r\^ole
of $M$ in the low-energy expansion and in the renormalization
of the whole effective theory. We
have calculated the one-loop contribution to all the
above mentioned quantities. Qualitatively, they do not
modify any of our conclusions.\\

{\bf Acknowledgments} 
This work was supported in part by 
EC-Contract
No. ERBFMRX-CT980169 (EURODA$\Phi$NE).

\end{document}